\documentclass{article}

\usepackage{graphicx}
\usepackage{subcaption}
\usepackage{hyperref}

\usepackage{authblk}

\usepackage[final,nonatbib]{neurips_2018}

\usepackage[utf8]{inputenc} 
\usepackage[T1]{fontenc}    
\usepackage{hyperref}       
\usepackage{url}            
\usepackage{booktabs}       
\usepackage{amsfonts}       
\usepackage{nicefrac}       
\usepackage{microtype}      
\usepackage[numbers,sort]{natbib}

\title{A Case for the Score: Identifying Image Anomalies using Variational Autoencoder Gradients}

\author{
  \textbf{David Zimmerer, Jens Petersen, Simon A. A. Kohl} and \textbf{Klaus H. Maier-Hein} \\
  Division of Medical Image Computing\\
  German Cancer Research Center (DKFZ)\\
  Heidelberg, Germany \\
  \texttt{\{d.zimmerer,jens.petersen,simon.kohl,k.maier-hein\}@dkfz.de}
}

\begin{document}

\maketitle

\begin{abstract}
  Through training on unlabeled data, anomaly detection has the potential to impact computer-aided diagnosis by outlining suspicious regions. 
  Previous work on deep-learning-based anomaly detection has primarily focused on the reconstruction error. We argue instead, that pixel-wise anomaly ratings derived from a Variational Autoencoder based \textit{score} approximation yield a theoretically better grounded and more faithful estimate.
  In our experiments, Variational Autoencoder gradient-based rating outperforms other approaches on unsupervised pixel-wise tumor detection on the BraTS-2017 
  dataset with a ROC-AUC of 0.94.
\end{abstract}

\section{Introduction}

In recent years several deep-learning-based methods have reported reaching comparable performance to trained medical physicians \cite{liu_detecting_2017,gulshan_development_2106}.  
One weakness of those approaches is that they still require a lot of annotated data for each condition to be trained on.
Due to the time-intensive work of annotating medical images and the combinatorial number of cases for different modalities, image qualities, hardware devices, and different conditions, it is still infeasible to train an algorithm for each of the existing combinations.
Anomaly detection can, while not determining the condition, highlight and identify suspicious regions for a closer inspection by a trained physician.
By assigning each pixel an anomaly rating, it allows for an easy trade-off of specificity and sensitivity. 
While this may not be able to outperform supervised algorithms, it offers a way to make use of unlabeled data and aid physicians during the diagnosis.

Previous unsupervised anomaly detection approaches in the medical field were primarily based on a reconstruction error.
Leemput et al. \cite{van_leemput_automated_2001} use a statistical model to reconstruct the input tissue-wise, quantifying the discrepancies between the actual image and the model prediction to identify anomalies.
Liu et al. \cite{liu_low-rank_2014} decompose the model into low-rank components which representing the normal parts of the image, and high-frequency parts which representing anatomical and pathological variations and are thus able to delineate suspicious areas.
More recently multiple deep learning Autoencoder (AE) based methods have been proposed, all considering the reconstruction error.
Chen et al. \cite{chen_unsupervised_2018,chen_deep_2018} propose to use an adversarial latent loss in addition to a Variational Autoencoder (VAE) and compare it to different AE-based approaches. 
Baur et al. \cite{baur_deep_2018} use a VAE with an adversarial loss on the reconstruction to get a more realistic reconstruction.
Pawlowski et al. \cite{pawlowski_unsupervised_2018} compare different AEs for CT based pixel-wise segmentation.

All those approaches use the reconstruction error to identify suspicious regions, based on the idea that models can not truthfully reproduce anomalies not seen during training. 
Despite showing good results, there are no formal guarantees for that assumption. 
In the next section we will describe how to use the \textit{score}, defined as the derivative of the log-density with respect to the input $\frac{\partial \log p(x)}{\partial x}$  \cite{hyvarinen_estimation_2005}, as an alternative anomaly rating.



\section{Methods}

Alain et al. \cite{alain_what_2014} have shown that for AE-based models with a denoising criterion the reconstruction error approximates the \textit{score}.
It can be anticipated that most AE- and reconstruction-based models work due to an approximation of the \textit{score}.
Consequently and based on the following assumptions, we hypothesize that the \textit{score} can give a good approximation for an abnormality rating:
\begin{itemize}
    \item The \textit{score} gives the directions towards the normal data samples, which for medical data is the data sample with abnormal anatomies and pathologies transformed into healthy parts,
    \item The magnitude of the \textit{score} indicates how abnormal the pixel is.
\end{itemize}

In this work, we describe a way to directly estimate the \textit{score} using VAEs, one of the best performing density-estimation models for images \cite{chen_deep_2018,kiran_overview_2018}.
The objective of VAEs is to learn a generative model of the data by maximizing the evidence lower bound (ELBO) for the given training data. 
The ELBO is defined as:
\begin{equation}
\label{eq:elbo}
\log p(x) \geq -D_{KL}(q(z|x) || p(z)) + \mathbb{E}_{q(z|x)} [\log p(x|z)],
\end{equation}
Where $q(z|x)$ is the inference model, $p(z)$ is the prior for the latent variables, $D_{KL}$ is the Kullback-Leibler divergence, and $p(x|z)$ is the generative model.
Thus after training the VAE and maximizing the ELBO, an estimate of the log probability $\log p(x)$ of a data sample $x$ can be calculated by evaluating the rhs of Eq. \ref{eq:elbo} for the data sample $x$.
The approximate \textit{score} can consequently be calculated by taking the derivative of the ELBO with respect to the data sample:
\begin{equation}
\label{eq:vaescore}
\frac{\partial \log p(x)}{\partial x} \approx \frac{\partial (-D_{KL}(q(z|x) || p(z)) + \mathbb{E}_{q(z|x)} [\log p(x|z)])}{\partial x},
\end{equation}
Furthermore, the ELBO is fully differentiable \cite{kingma_auto-encoding_2013,rezende_stochastic_2014}, when training a VAE using Gaussian distributions for $p(z)$ and $p(x|z)$, a parameterization by neural networks, the reparameterization trick, and MC sampling to approximate the expectation.
This allows training of the VAE and the evaluation of Eq. \ref{eq:vaescore} using the backpropagation algorithm.

We note that the above-mentioned assumptions
can be violated in practice, especially in cases far away from the healthy sample data distribution. However, in the next section, we will present empirical evidence that our model can outperform reconstruction-based methods on an anomaly detection tasks and describe its benefits.


\section{Experiments \& Results}

To learn the healthy data distribution we trained the VAE model on 1092 T2 MRI images of Human Connectome Project (HCP) dataset \cite{van_essen_human_2012}, with minor data augmentations, such as multiplicative color augmentations, random mirroring, and rotations.
We evaluate the anomaly detection in the context of finding and outlining tumors on the BraTS-2107 dataset \cite{bakas_advancing_2017,menze_multimodal_2015}.
Therefore we calculate a pixel-wise rating and then report the ROC-AUC. 
Both datasets were normalized and slice-wise resampled to a resolution of 64x64 pixels. 
As encoder and decoder for the AE-based models, we used a 5-layer fully convolutional neural network with LeakyReLUs and a latent size of 1024. 
To backpropagate onto the image and approximate the \textit{score}, we used the Smoothgrad algorithm \cite{smilkov_smoothgrad_2017}. Due to checkerboard artifacts caused by the convolutions, we apply Gaussian smoothing to the gradients. 
The model was trained for 60 epochs with a batchsize of 64 and Adam as the optimizer with a learning rate of $0.0002$.

\begin{figure}[ht!]
  \centering
  \begin{subfigure}[b]{0.95\textwidth}
      \centering
      \includegraphics[width=0.95\textwidth]{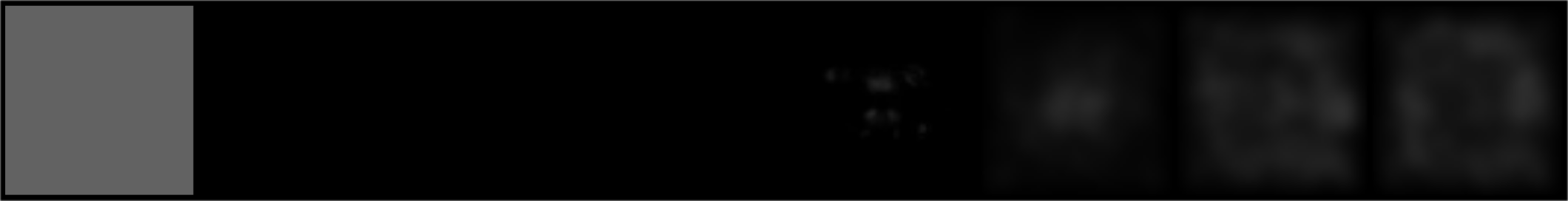}
  \end{subfigure}
  \begin{subfigure}[b]{0.95\textwidth}
      \centering
      \includegraphics[width=0.95\textwidth]{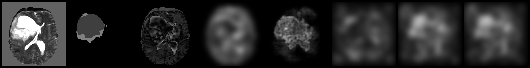}
  \end{subfigure}
  \begin{subfigure}[b]{0.95\textwidth}
      \centering
      \includegraphics[width=0.95\textwidth]{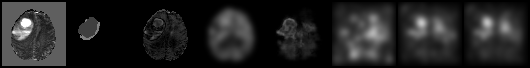}
  \end{subfigure}
  \begin{subfigure}[b]{0.95\textwidth}
      \centering
      \includegraphics[width=0.95\textwidth]{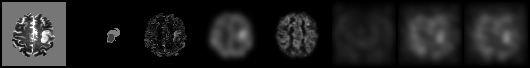}
  \end{subfigure}
  \begin{subfigure}[b]{0.95\textwidth}
      \centering
      \includegraphics[width=0.95\textwidth]{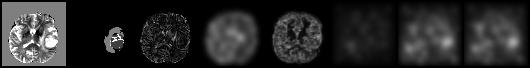}
  \end{subfigure}
  \begin{subfigure}[b]{0.95\textwidth}
      \centering
      \includegraphics[width=0.95\textwidth]{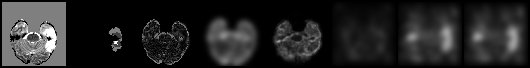}
  \end{subfigure}
  \begin{subfigure}[b]{0.95\textwidth}
      \centering
      \includegraphics[width=0.95\textwidth]{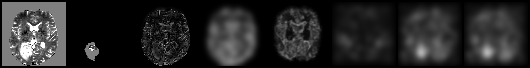}
  \end{subfigure}
  \begin{subfigure}[b]{0.95\textwidth}
      \centering
      \includegraphics[width=0.95\textwidth]{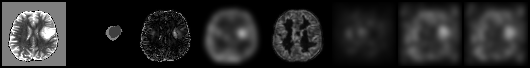}
  \end{subfigure}
  \begin{subfigure}[b]{0.95\textwidth}
      \centering
      \includegraphics[width=0.95\textwidth]{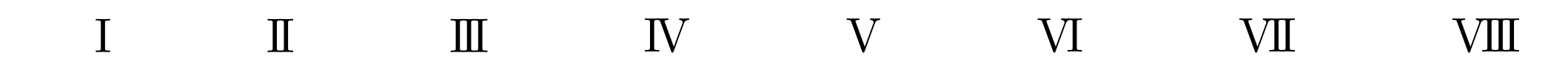}
  \end{subfigure}
  
\caption{More samples as presented in Fig. \ref{fig:samples}, showing the original sample (I), the annotation (II), the reconstruction error (III), the smoothed reconstruction error (IV), the sampling variances (V), the reconstruction-loss gradient  (VI), the KL-loss gradient (VII), and the ELBO gradient which approximates the \textit{score} (VIII).}
\label{fig:more1}
\end{figure}
\begin{figure}[ht!]
  \begin{subfigure}[b]{0.95\textwidth}
      \centering
      \includegraphics[width=0.95\textwidth]{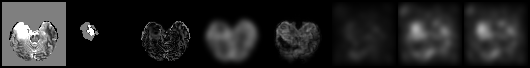}
  \end{subfigure}
  \begin{subfigure}[b]{0.95\textwidth}
      \centering
      \includegraphics[width=0.95\textwidth]{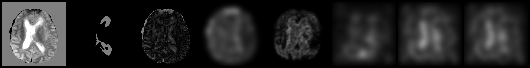}
  \end{subfigure}
  \begin{subfigure}[b]{0.95\textwidth}
      \centering
      \includegraphics[width=0.95\textwidth]{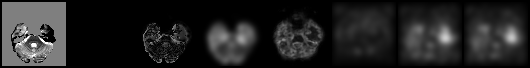}
  \end{subfigure}
  \begin{subfigure}[b]{0.95\textwidth}
      \centering
      \includegraphics[width=0.95\textwidth]{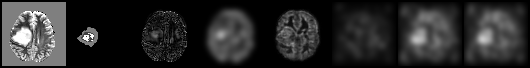}
  \end{subfigure}
  \begin{subfigure}[b]{0.95\textwidth}
      \centering
      \includegraphics[width=0.95\textwidth]{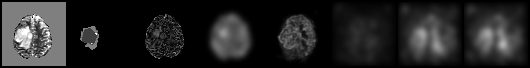}
  \end{subfigure}
  \begin{subfigure}[b]{0.95\textwidth}
      \centering
      \includegraphics[width=0.95\textwidth]{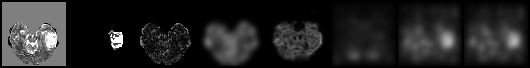}
  \end{subfigure}
  \begin{subfigure}[b]{0.95\textwidth}
      \centering
      \includegraphics[width=0.95\textwidth]{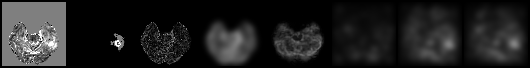}
  \end{subfigure}
  \begin{subfigure}[b]{0.95\textwidth}
      \centering
      \includegraphics[width=0.95\textwidth]{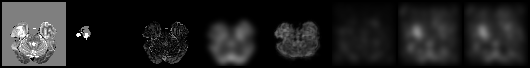}
  \end{subfigure}
  \begin{subfigure}[b]{0.95\textwidth}
      \centering
      \includegraphics[width=0.95\textwidth]{samples/grid.png}
  \end{subfigure}

\caption{More samples as presented in Fig. \ref{fig:samples}, showing the original sample (I), the annotation (II), the reconstruction error (III), the smoothed reconstruction error (IV), the sampling variances (V), the reconstruction-loss gradient  (VI), the KL-loss gradient (VII), and the ELBO gradient which approximates the \textit{score} (VIII).}
\label{fig:more2}
\end{figure}

To evaluate the benefits of the \textit{score}, we compare the model to a Denoising Autoencoder (DAE) \cite{vincent_stacked_2010} with the same architecture using the reconstruction error.
Furthermore, we compare the \textit{score} with the reconstruction error of the VAE, the smoothed reconstruction error, and the sampling deviations by determining the standard deviation of multiple MC samples.
We further inspect the \textit{score}, dividing it into the reconstruction-loss gradient and KL-loss gradient to get insights into the benefits of including the KL-term into the anomaly detection.
The results can be seen in Fig. \ref{fig:comparison}a (and Appendix Table \ref{tab:results}), samples and the corresponding pixel-wise ratings for samples are presented in Fig. \ref{fig:samples}b (and Appendix Fig. \ref{fig:more1} \& \ref{fig:more2}).
 
The reconstruction error performs similarly for the VAE and the DAE, which was also reported in \cite{chen_deep_2018,pawlowski_unsupervised_2018}. Smoothing leads to slightly improved results, presumably by removing high-frequency detections, and performs on par with the usage of the sampling variances.
The approximated \textit{score} using the ELBO gradient (KL-loss + reconstruction-loss) performs best with a pixel-wise ROC-AUC of 0.94 (see Appendix Fig. \ref{fig:curves}) .
It is interesting to see, that the addition of the reconstruction-loss to the KL-loss shows little benefit over the KL-loss gradient. Furthermore, the reconstruction-loss gradient performs worse than the KL-loss gradient but outperforms the reconstruction error.

In Fig. \ref{fig:samples}a, the reconstruction-loss gradient focuses on parts of poor reconstruction, and the combination of the KL-loss with the reconstruction-loss shows only marginal benefits over the KL-loss gradient. 
This might be an indication that for this model the KL-loss focuses primarily on the distance to the data distribution, while the reconstruction focuses more on the actual reconstruction task.

\begin{table}[tb]

\begin{minipage}[c]{0.05\linewidth}
(a)
\end{minipage}
\hspace{0.5cm}
\begin{minipage}[c]{0.88\linewidth}
    \centering
          \includegraphics[width=0.7\textwidth]{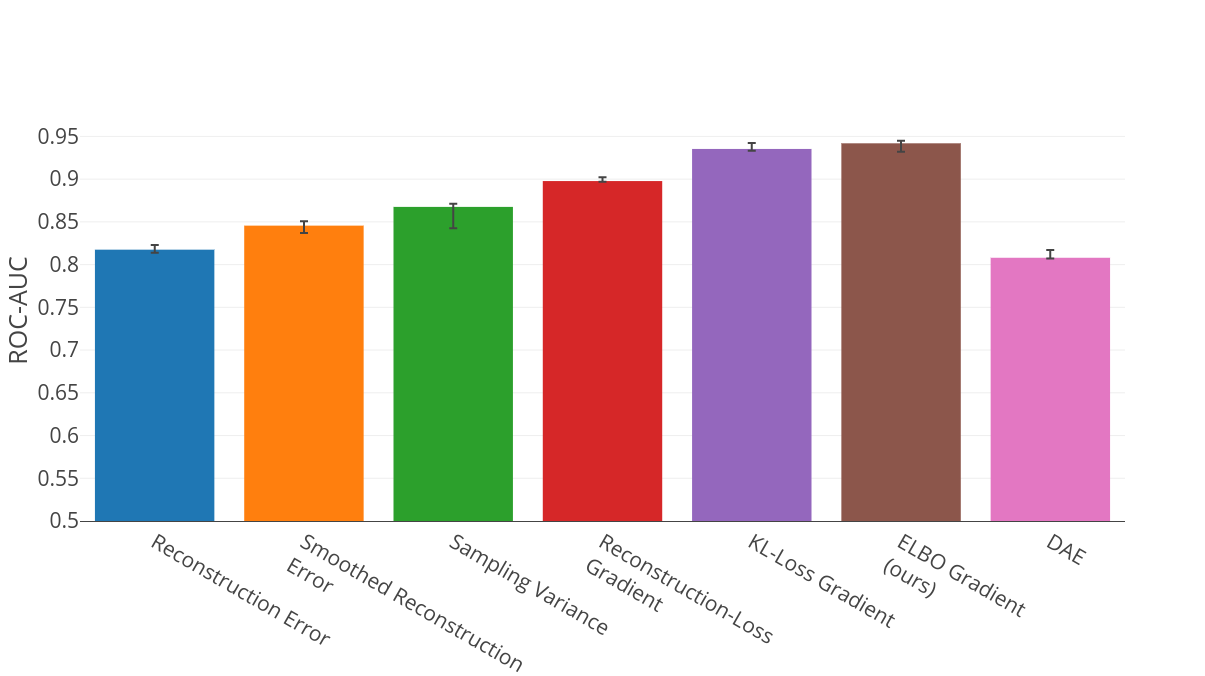}
\end{minipage}

\begin{minipage}[c]{0.05\linewidth}
(b)
\end{minipage}
\hspace{0.5cm}
\begin{minipage}[c]{0.84\linewidth}
\vspace{1.0em}

    \centering
  \begin{subfigure}[b]{\textwidth}
    \includegraphics[width=0.95\textwidth]{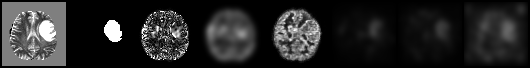}
  \end{subfigure}
  \begin{subfigure}[b]{\textwidth}
    \includegraphics[width=0.95\textwidth]{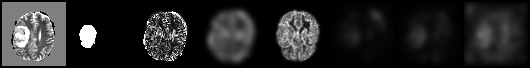}
  \end{subfigure}
  \begin{subfigure}[b]{\textwidth}
    \includegraphics[width=0.95\textwidth]{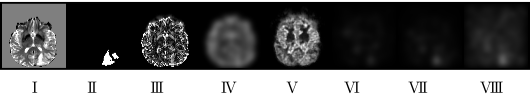}
  \end{subfigure}
\end{minipage}

\captionof{figure}{(a) Comparison of the pixel-wise tumor detection ROC-AUC on the BraTS-2017 dataset.
(b) Samples from the dataset with the different pixel-wise rating schemes, showing the original sample (I), the annotation (II), the reconstruction error (III), the smoothed reconstruction error (IV), the sampling variances (V), the reconstruction-loss gradient  (VI), the KL-loss gradient (VII), and the ELBO gradient which approximates the \textit{score} (VIII).
}
\label{fig:samples}
\label{fig:comparison}
\end{table}

\subsection{Discussion \& Conclusion}

We have presented a way to estimate the \textit{score} using VAE gradients to detect anomalies on the BraTS-2017 tumor segmentation dataset. The results show competitive unsupervised segmentation performance, slightly outperforming the previously best reported ROC-AUC of 0.92 \cite{chen_unsupervised_2018,chen_deep_2018}.
The relative influence of the reconstruction loss can depend on the regularization of the latent variables. Using fewer latent variables or putting more importance on the KL-loss could, while potentially causing inferior overall performance, lead to a more competitive performance of the reconstruction error.

To the best of our knowledge, we are the first to use the gradients of a VAE, which approximate the  \textit{score}, to identify anomalies in images. 
The results suggest that the approximated \textit{score}, including the often ignored KL-loss, can give a boost on the pixel-wise anomaly detection performance. Furthermore, we want to stress the point that including the KL-loss for a pixel-wise anomaly detection and the \textit{score} of a model can lead to an improvement in VAE-based methods for pixel-wise anomaly ratings.

This method should also be directly applicable to other state-of-the-art density estimation techniques, such as Grow \cite{kingma_glow:_2018} or Pixel-CNN++ \cite{salimans_pixel_2017}, and it would be an interesting next step to see how different models perform.




\bibliography{refs}
\bibliographystyle{abbrv}

\newpage

\section{Appendix}

\subsection{Quantitative Results}

\begin{table}[ht!]
\renewcommand\thetable{1}
\centering
\begin{tabular}{|l|l|}
\hline
                              & ROC-AUC        \\ \hline
DAE                           & $0.808 \pm 0.009$ \\ \hline
Reconstruction Error          & $0.817 \pm 0.003$ \\ \hline
Smoothed Reconstruction Error & $0.843 \pm 0.008$ \\ \hline
Sampling Variance             & $0.855 \pm 0.013$ \\ \hline
Reconstruction-Loss Gradient  & $0.894 \pm 0.020$ \\ \hline
KL-Loss Gradient              & $0.939 \pm 0.007$ \\ \hline
ELBO Gradient                 & $0.939 \pm 0.008$ \\ \hline
\end{tabular}
\vspace{0.7em}
\caption{Pixel-wise ROC-AUC values of the compared approaches (see Fig. \ref{fig:comparison}).}
\label{tab:results}
\end{table}

\begin{figure}[ht!]
  \centering
  \begin{subfigure}[b]{0.49\textwidth}
      \centering
      \includegraphics[width=0.95\textwidth]{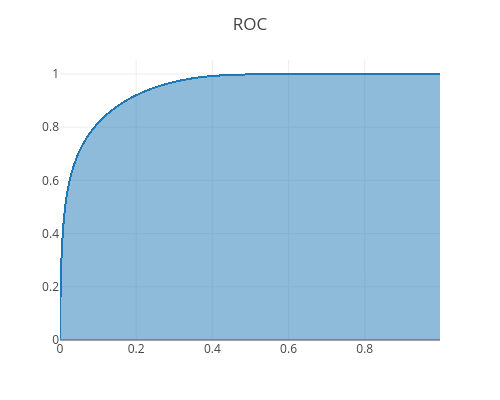}
  \end{subfigure}
  \begin{subfigure}[b]{0.49\textwidth}
    \centering
    \includegraphics[width=0.95\textwidth]{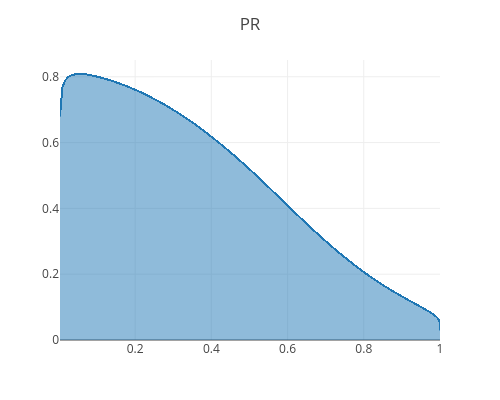}
   \end{subfigure}
\caption{Pixel-wise Reciver Operator Curve (ROC) and Precision Recall (PR) Curve on the test set for the VAE ELBO-gradient with regard to the anomaly labels (all annotations are considered anomalies).}
\label{fig:curves}
\end{figure}

\newpage

\subsection{Qualitative Results}

\end{document}